\documentclass[preprint,12pt, a4paper]{elsarticle}

\usepackage{amssymb}
\usepackage[hyphens,spaces,obeyspaces]{url}
\usepackage[colorlinks=true,allcolors=blue]{hyperref}
\usepackage{subcaption}
\usepackage{adjustbox}
\usepackage{caption}
\usepackage{floatrow}

\usepackage{float}

\usepackage{xcolor}

\usepackage{listings}
\definecolor{deepblue}{rgb}{0,0,0.5}
\definecolor{deepred}{rgb}{0.6,0,0}
\definecolor{deepgreen}{rgb}{0,0.5,0}
\DeclareFixedFont{\ttb}{T1}{txtt}{bx}{n}{8} 
\DeclareFixedFont{\ttm}{T1}{txtt}{m}{n}{8}  
\newfloat{Listing}{hbtp}{lop}

\restylefloat{table}
\journal{SoftwareX}

\begin{document}

\begin{frontmatter}



\title{Enabling MPI communication within Numba/LLVM JIT-compiled Python code using {\bf numba-mpi} v1.0}


\author[UJ]{Kacper~Derlatka\footnote{former student, currently at pega.com}}
\author[UJ]{Maciej~Manna\footnote{former student, currently at autodesk.com}}
\author[UJ]{Oleksii~Bulenok\footnote{former student, currently at swmansion.com}}
\author[MPIDS]{David~Zwicker}
\author[AGH]{Sylwester~Arabas}

\address[UJ]{Faculty of Mathematics and Computer Science, Jagiellonian University, Kraków, Poland\!\!\!}
\address[MPIDS]{Max Planck Institute for Dynamics and Self-Organization, Göttingen, Germany}
\address[AGH]{Faculty of Physics and Applied Computer Science, AGH Univeristy of Krakow, Poland}

\begin{abstract}
The \verb=numba-mpi= package offers access to the Message Passing Interface (MPI) routines 
  from Python code that uses the Numba just-in-time (JIT) compiler. 
As a result, high-performance and multi-threaded Python code may utilize MPI 
  communication facilities without leaving the JIT-compiled code blocks,
  which is not possible with the \verb=mpi4py= package, a higher-level Python interface to MPI. 
For debugging purposes, \verb=numba-mpi= retains full functionality of the code even if the 
  JIT compilation is disabled. 
The \verb=numba-mpi= API constitutes a thin wrapper around the C API of MPI and is built around Numpy arrays including handling 
  of non-contiguous views over array slices. 
Project development is hosted at GitHub leveraging the \verb=mpi4py/setup-mpi= workflow enabling continuous integration 
  tests on Linux (\verb=MPICH=, \verb=OpenMPI= \& \verb=Intel MPI=), macOS (\verb=MPICH= \& \verb=OpenMPI=) and Windows (\verb=MS MPI=). 
The paper covers an overview of the package features, architecture and performance.
As of v1.0, the following MPI routines are exposed and covered by unit tests: \verb=size=/\verb=rank=, \verb=[i]send=/\verb=[i]recv=,
  \verb=wait[all|any]=, \verb=test[all|any]=, \verb=allreduce=, \verb=bcast=, \verb=barrier=, \verb=scatter/[all]gather= \& \verb=wtime=. 
The package is implemented in pure Python and depends on \verb=numpy=, \verb=numba= and \verb=mpi4py= (the latter used 
  at initialization and as a source of utility routines only).
The performance advantage of using \verb=numba-mpi= compared to \verb=mpi4py= is depicted with 
  a simple example, with entirety of the code included in listings discussed in the text.
Application of \verb=numba-mpi= for handling domain decomposition in numerical solvers for partial differential 
  equations is presented using two external packages that depend on \verb=numba-mpi=: \verb=py-pde= and \verb=PyMPDATA-MPI=.
\end{abstract}

        
\end{frontmatter}


\section*{Current code version}

\begin{table}[H]\footnotesize
\caption{Code metadata.}
\begin{tabular}{|p{.4cm}|p{5.3cm}|p{6.8cm}|}
\hline
\textbf{Nr.} & \textbf{Code metadata description} & \\
\hline
C1 & Current code version & 1.0.0 \\
\hline
C2 & Permanent link to code/repository used for this code version & \href{https://doi.org/10.5281/zenodo.12609957}{DOI:10.5281/zenodo.12609957} \\
\hline
C3 & Code Ocean compute capsule & None \\
\hline
C4 & Legal Code License & GNU GPL v3.0 \\
\hline
C5 & Code versioning system used & git (\href{https://github.com/numba-mpi/numba-mpi}{github.com/numba-mpi/numba-mpi})\\
\hline
C6 & Software code languages, tools, and services used & Python, MPI \\
\hline
C7 & Compilation requirements, operating environments \& dependencies & Dependencies: \verb=mpi4py=, \verb=numpy=, \verb=psutil= 
\newline OS: Linux, macOS, Windows\\
\hline
C8 & Link to developer documentation/manual & \href{https://numba-mpi.github.io/numba-mpi}{numba-mpi.github.io/numba-mpi} \\
\hline
C9 & Support email for questions & please use GitHub issue tracker \\
\hline
\end{tabular}
\label{} 
\end{table}


\section*{Introduction}

The technologies around which the present work revolves are:
(i) Python programming language, (ii) Message Passing Interface (MPI) - the de-facto standard parallel supercomputing communication framework and
(iii) Numba.
Numba \cite{Numba_Lam_et_al_2015} bridges the gap between Python and high-performance computing by offering just-in-time (JIT) compilation, multi-threading parallelism and GPU computing.
The aim of the hereby introduced \verb=numba-mpi= package is to extend Numba  by making it interoperable with the MPI communication model.
The paper is structured as follows:
\begin{description}
    \vspace{-.7em}
    \item[Section \ref{sec:prereq}]{explains why combining Numba with MPI offers significant potential for high-performance parallel computing in Python, and explains why it had not been possible without \verb=numba-mpi=;}
    
    \vspace{-.7em}
    \item[Section \ref{sec:impl}]{outlines the design and implementation, package availability, test coverage and testing workflows;}
        
    \vspace{-.7em}
    \item[Section \ref{sec:examples}]{describes two examples of external Python packages for solving partial differential equations (PDEs) that rely on \verb=numba-mpi= for handling domain decomposition;}

    \vspace{-.7em}
    \item[Section \ref{sec:concl}]{serves as a conclusion that highlights  key achievements so far, enumerates remaining developments for which we welcome community contributions, and outlines the potential impact of \verb=numba-mpi=.}
\end{description}
    
\section{Background and motivation}\label{sec:prereq}

Python and MPI share development timelines --- both had their 1.0 releases in early 1990s, 2.0 and 3.0 following roughly a decade apart\footnote{
See \url{https://docs.python.org/3/license.html\#history-of-the-software}, \url{https://devguide.python.org/versions/} and \url{https://www.mpi-forum.org/docs/}
}.
Both technologies have been widely adopted in scientific computing, yet in somewhat antipodal roles: Python as an interpreted, inherently-serial, high-level language commonly used in interactive contexts; MPI as a low/middle-level communication/synchronization paradigm requiring custom compilation mechanisms and used in massively parallel applications.
Both became de-facto standards in their respective domains, namely: Python in scientific computing (as a language-of-choice for introductory courses \cite{Landau_et_al_2024} and a sough-after transversal skill \cite{StckOverflow2023}) and MPI in high-performance massively-parallel computing, both in academia and industry \cite{Osseyran_and_Giles_2015,Sterling_et_al_2018}).
MPI also finds its niche in high-performance cloud-computing applications \cite{Xu_et_al_2020}.

\subsection{Accelerating Python code with Numba}

There have been several solutions developed that bring together Python and high-performance computing (see \cite{Castro_et_al_2023} for a recent review). These include static compilation solutions (e.g., Cython \cite{Behnel_et_al_2011}, Pythran \cite{Guelton_2018}) and JIT-based tools such as PyPy \cite{Bolz_et_al_2009} (engineered as a drop-in replacement for the standard CPython interpreter, capable of handling the entirety of SciPy test suite \cite{Virtaen_et_al_2020}), JAX \cite{Frostig_et_al_2018} and Numba \cite{Numba_Lam_et_al_2015} (both engineered as a Python packages for use with the CPython interpreter). Numba is built on top the LLVM Compiler Infrastructure \cite{Lattner_et_al_2004} and specifically targets acceleration of codes featuring numerical algorithms. Besides JIT compilation, Numba offers high-level OpenMP-like multi-threading facility, as well as GPU programming API supporting NVIDIA CUDA.

Listing~\ref{lst:hello_world} includes a code snippet and its output illustrating the way one uses Numba and the speedup potential it offers. 
The snippet includes definition of a function \verb=get_pi_part()= estimating the value of $\pi$ by numerical integration of $\int_0^1 (4/(1+x^2))\mathrm{d}x=\pi$. 
The function performs the calculation by dividing the integration interval $(0,1)$  into \verb=n_intervals= sub-intervals and determining the associated Riemann sum (operation with non-default arguments will be discussed in section \ref{sec:numba_mpi_code_example} below). 
In line~3, the \verb=@numba.jit= decorator is used instructing Numba to compile the function into machine code (\verb;nopython=True; opts out from  any fallback involvement of the Python interpreter ensuring that entirety of the decorated function can be JIT-compiled to machine code).
In lines~12--13, a comparison is made between times it took to execute the function without (\verb=get_pi_part.py_func()=) and with the compilation enabled.
The reported speedup is of the order of 100.
The example depicts the key benefits of using Numba, namely: the significant performance gain compared to execution with standard CPython; minimal modification to the code needed to obtain the speedup; the option to execute the code with JIT compilation disabled for debugging or benchmarking purposes.

\begin{Listing}[t]
  {\bf\textsf{script:}}
\begin{lstlisting}
import numba, timeit

@numba.jit(nopython=True)
def get_pi_part(n_intervals=1000000, rank=0, size=1):
    h = 1 / n_intervals
    partial_sum = 0.0
    for i in range(rank + 1, n_intervals, size):
        x = h * (i - 0.5)
        partial_sum += 4 / (1 + x**2)
    return h * partial_sum

time = lambda fun: min(timeit.repeat(fun, number=1, repeat=5))
print(f"speedup: {time(get_pi_part.py_func) / time(get_pi_part):.3g}")
\end{lstlisting}
  {\bf\textsf{output:}}\\
  \colorbox{black}{\begin{minipage}{.98\textwidth}\color{white}\bf\texttt{%
speedup: 97.5 
  }\end{minipage}}
  \vspace{-.75em}
  \caption{\label{lst:hello_world}
    Definition of \texttt{get\_pi\_part()} function and a comparison of its performance with Numba JIT enabled and disabled (result obtained with Python 3.12, and Numba 0.60.0)
  }
\end{Listing}

\subsection{Using MPI from Python with mpi4py and its incompatibility with Numba}

The ``toy example'' algorithm used above is readily parallelisable and will serve hereinafter to showcase how MPI communication routines can be used from Python.
The parallelization strategy leverages independence of computations within individual intervals, and relies on MPI reduction functionality for handling accumulation of partial sums computed by separate processes (serving here as an ersatz of supercomputing nodes).

The leading solution for using MPI from Python is \verb=mpi4py= \cite{Dalcin_and_Fang_2021,Fink_et_al_2021} which is implemented in Cython.
Code in Listing~\ref{lst:error} defines a function \verb=pi_mpi4py()=, which orchestrates parallel execution of \verb=get_pi_part()= over subsets of intervals and subsequent summation of the results using MPI.
Repeating the computations \verb=N_TIMES= is done solely to make timing meaningful.
In lines~16--19, an attempt is made to JIT-compile the newly defined function,
which fails with the error message given below the listing and depicting the issue of incompatibility of Numba with \verb=mpi4py=.
Let us underline here that function \verb=pi_mpi4py()= is correct and can be used to perform its task still benefiting from Numba acceleration of \verb=get_pi_part()=. However, JIT-compilation of the MPI communication logic based on \verb=mpi4py= is impossible.
Consequently, there is a need to perform ``roundtrips'' between the JIT-compiled and the interpreted code in each iteration (cost of which is depicted in section~\ref{sec:perfgain} herein).
The aim of \verb=numba-mpi= is to eliminate these ``roundtrips''.

\begin{Listing}[t]
  {\bf\textsf{script (depends on code from Listing~\ref{lst:hello_world}):}}
\begin{lstlisting}
import mpi4py, numba, numpy as np

N_TIMES = 10000
RTOL = 1e-3

def pi_mpi4py(n_intervals):
    pi = np.array([0.])
    part = np.empty_like(pi)
    for _ in range(N_TIMES):
        part[0] = get_pi_part(n_intervals,
            mpi4py.MPI.COMM_WORLD.rank, mpi4py.MPI.COMM_WORLD.size)
        mpi4py.MPI.COMM_WORLD.Allreduce(part,
            (pi, mpi4py.MPI.DOUBLE), op=mpi4py.MPI.SUM)
        assert abs(pi[0] - np.pi) / np.pi < RTOL

try:
    numba.jit(pi_mpi4py, nopython=True)(1)
except Exception as ex:
    print(ex)
\end{lstlisting}
  {\bf\textsf{output:}}\\
  \colorbox{black}{\begin{minipage}{.98\textwidth}\color{white}\bf\texttt{%
Failed in nopython mode pipeline (step: nopython frontend)
Internal error at resolving type of attribute "MPI" ...
  }\end{minipage}}
  \vspace{-.25em}
  \caption{\label{lst:error}
  Definition of \texttt{pi\_mpi4py} function orchestrating parallel processing using the \texttt{get\_pi\_part} defined in Listing~\ref{lst:hello_world} and using \texttt{mpi4py} for performing reduction operation across communicating workers. Below the definition, an attempt to JIT-compile \texttt{pi\_mpi4py} is included along with the error message printed in the \texttt{except} block (\texttt{mpi4py} 3.1.6).
  }
\end{Listing}

\section{Software description, usage basics and performance}\label{sec:impl}

\subsection{Using MPI from Python with numba-mpi}\label{sec:numba_mpi_code_example}

\begin{Listing}[t]
  {\bf\textsf{script (depends on code from Listings~\ref{lst:hello_world}-\ref{lst:error}):}}
\begin{lstlisting}
import numba_mpi

@numba.jit(nopython=True)
def pi_numba_mpi(n_intervals):
    pi = np.array([0.])
    part = np.empty_like(pi)
    for _ in range(N_TIMES):
        part[0] = get_pi_part(n_intervals, numba_mpi.rank(), numba_mpi.size())
        numba_mpi.allreduce(part, pi, numba_mpi.Operator.SUM)
        assert abs(pi[0] - np.pi) / np.pi < RTOL
\end{lstlisting}
  \caption{\label{lst:numba_mpi}
    Definition of \texttt{pi\_numba\_mpi} function performing analogous parallel processing logic as \texttt{pi\_mpi4py} but using \texttt{numba-mpi} instead of \texttt{mpi4py} and hence being susceptible to JIT compilation (and thus decorated with the \texttt{@numba.jit} decorator).
  }
\end{Listing}

In order to circumvent the interoperability issue demonstrated above, in this case preventing from using MPI via \verb=mpi4py= from within code JIT-compiled with Numba, \verb=numba-mpi= streamlines access to the C API of MPI using Python's built-in \verb=ctypes= foreign function library.
Numba supports \verb=ctypes=-exposed routines out of the box, and this route is used, e.g., for offering parts of the SciPy functionality to Numba JIT-compiled code in the \verb=numba-kdtree= package\footnote{\texttt{numba-kdtree} Python package: \url{https://pypi.org/p/numba-kdtree/}}.
Developing \verb=numba-mpi=, it was further decided to include \verb=mpi4py= as its dependency for handling loading of MPI dynamic library and thus not only simplifying the implementation but also ensuring that both \verb=mpi4py= and \verb=numba-mpi= can be used together.

The API of \verb=numba-mpi= closely resembles the C API of MPI, hence
constitutes a lower-level interface than \verb=mpi4py=.
Listing~\ref{lst:numba_mpi} depicts definition of a \verb=pi_numba_mpi()= function containing equivalent logic to \verb=pi_mpi4py()= from Listing~\ref{lst:error} with calls to \verb=mpi4py= API replaced with calls to \verb=numba-mpi= wrappers over the \verb=MPI C= API.
This enables the use of the \verb=@numba.jit= decorator and results in inclusion of all computations and communications within one JIT-compiled block of code avoiding repeated (\verb=N_TIMES= in this case) ``roundtrips'' between interpreted and compiled code just for the purpose of performing MPI communication.

\begin{Listing}[t]
  {\bf\textsf{script (depends on code from Listings~\ref{lst:hello_world}-\ref{lst:numba_mpi})}}
\begin{lstlisting}
from matplotlib import pyplot

plot_x = [x for x in range(1, 11)]
plot_y = {'numba_mpi': [], 'mpi4py': []}
for x in plot_x:
    for impl in plot_y:
        plot_y[impl].append(min(timeit.repeat(
            f"pi_{impl}(n_intervals={N_TIMES // x})",
            globals=locals(),
            number=1,
            repeat=10
        )))

if numba_mpi.rank() == 0:
    pyplot.plot(
        plot_x,
        np.array(plot_y['mpi4py'])/np.array(plot_y['numba_mpi']),
        marker='o'
    )
\end{lstlisting}
  \caption{\label{lst:timing}
    Python script utilizing definitions from Listings~\ref{lst:hello_world}-\ref{lst:numba_mpi}, performing execution time measurement and plotting resultant timings presented in Fig.~\ref{fig:perf}. Script intended to be executed via \texttt{mpiexec} to be run on multiple MPI workers (Fig.~\ref{fig:perf} presents measurements with work split across 4 workers).
  }
\end{Listing}

\subsection{Performance gain from using numba-mpi as compared to mpi}\label{sec:perfgain}

\begin{figure}
    \centering
    \includegraphics[width=\textwidth]{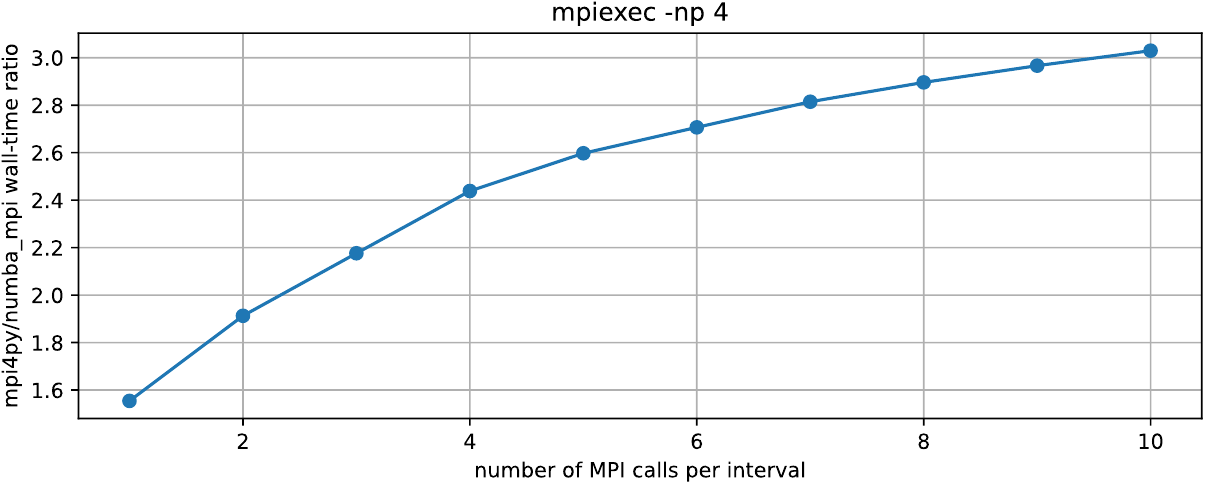}
    \vspace{-1.75em}
    \caption{Depiction of up to three-fold speedup obtained by using \texttt{numba-mpi} instead of \texttt{mpi4py} to avoid leaving the JIT-compiled code blocks. Figure created using the script in Listing~\ref{lst:timing} (and using code Listings~\ref{lst:hello_world}-\ref{lst:numba_mpi}).
    \label{fig:perf}
    \vspace{-.75em}
    }
\end{figure}

Listing~\ref{lst:timing} includes code that compares the performance of \verb=pi_numba_mpi()= versus that of \verb=pi_mpi4py()=.
Measurements are repeated \verb=N_REPEAT= times and the minimum value is reported.
The first execution of the JIT-compiled functions triggers compilation, hence selecting the minimum from several runs also serves to exclude the compilation time.
The resulting plot shown in Figure~\ref{fig:perf} depicts the speedup obtained by replacing \verb=mpi4py= with \verb=numba_mpi=, plotted as a function of \verb=N_TIMES / n_intervals= -- the number of MPI calls per interval.
The speedup, which stems from avoiding ``roundtrips'' between JIT-compiled and Python code is significant (150\%-300\%) in all cases.
The more often communication is needed (smaller \verb=n_intervals=), the larger the speedup.
Noteworthily, nothing in the actual number crunching (within the \verb=get_pi_part()= function) or in the employed communication logic (handled by the same MPI library) differs between the two compared solutions.
It is the overhead of repeatedly entering and leaving the JIT-compiled block if using \verb=mpi4py= 
that can be eliminated by using \verb=numba-mpi=.

\subsection{Philosophy and current status of the API}

The guiding principle of the public interface provided by \verb=numba-mpi= follows from its main use in high-performance computing.
For this reason, the API is quite minimal, low-level, and does not include any abstractions that would burden its users with extra costs.
The \verb=numba-mpi= provides a procedural interface with functions that should be familiar to users of MPI.
This is somewhat in contrast to the \verb=mpi4py= package that uses more object-oriented design that may be more convenient for some users, but does not surrender as easily to Numba JIT and may introduce extra performance costs.

Despite such low-level approach, \verb=numba-mpi= introduces some higher-level abstractions for the convenience of its users.
The signatures of MPI functions do not require supplying data types or sizes, as they can be easily deduced from the data itself, that is expected to be contained in NumPy arrays \cite{Harris2020}.
In case of some functions, \verb=numba-mpi= also handles data in arrays that are not necessarily contiguous \cite{Nolp_and_Oden_2023}.
An example of another convenience is automatic setup and cleanup of MPI runtime when package is imported.
The \verb=initialized= function may be used to check whether the initialization was successful.
Besides that, basic utility functions are provided, like \verb=size= and \verb=rank=, as well as \verb=wtime= for basic wall-clock timing of execution.
Moreover, the \verb=barrier= function is available for further synchronization capabilities.

\begin{Listing}[th!]
\begin{lstlisting}
import numpy as np
import numba_mpi as mpi
import numba

@numba.jit(nopython=True)
def exchange_data(src_data):
    dst_data = np.empty_like(src_data)
    reqs = np.zeros((2,), dtype=mpi.RequestType)

    if mpi.rank() == 0:
        status, reqs[0:1] = mpi.isend(src_data, dest=1, tag=11)
        assert status == mpi.SUCCESS
        status, reqs[1:2] = mpi.irecv(dst_data, source=1, tag=22)
        assert status == mpi.SUCCESS
    elif mpi.rank() == 1:
        status, reqs[0:1] = mpi.isend(src_data, dest=0, tag=22)
        assert status == mpi.SUCCESS
        status, reqs[1:2] = mpi.irecv(dst_data, source=0, tag=11)
        assert status == mpi.SUCCESS

    mpi.waitall(reqs)
    return dst_data

if mpi.rank() < 2:
    src_data = np.random.rand(10000)
    dst_data = exchange_data(src_data)
\end{lstlisting}
    \caption{\label{lst:exchange}
        Basic example of code used for non-blocking exchange of data between two processes using asynchronous operations \texttt{isend} and \texttt{irecv}.}
\end{Listing}

The core of the package includes synchronous data transfer functions - \verb=send=, \verb=recv=, \verb=bcast=, \verb=scatter=, \verb=gather=, \verb=allgather=, and \verb=allreduce=.
All of these functions use a signature that is as close as possible to the original interface of MPI.
In all cases, key parameters include the \verb=data= payload that is expected to be a NumPy array and integer identifier that specifies the rank of the distinguished process (\verb=dest= for \verb=send=, \verb=source= for \verb=recv=, \verb=root= for \verb=bcast=, etc.).
The \verb=tag= parameter is included where applicable, but its use is optional, and may be left out for convenience in simpler use cases. 
As already mentioned, \verb=type= and \verb=size= are omitted in the API since these parameters are deduced from the properties of \verb=data=.
For other function-specific parameters, like \verb=operator= used in \verb=allreduce=, all supported constants are provided (\verb=Operator= enumeration, with default value of \verb=SUM=).
A notable omission is the communicator (\verb=comm=) parameter that cannot be customized at this point and thus defaults to \verb=MPI_COMM_WORLD= (although support for this functionality is planned in near future).
All these functions return the status code -- either \verb=MPI_SUCCESS= or any code that indicates what errors did occur.

Finally, we provide a group of functions involving asynchronous operations -- \verb=isend= and \verb=irecv=.
They follow the same principles as described above, but they return -- besides the status code -- a handle to the \verb=Request= that can be used to query whether these operations are completed.
In particular, these handles (or arrays thereof) may be used with the special functions -- \verb=wait=, \verb=waitall=, \verb=waitany=, \verb=test=, \verb=testall=, and \verb=testany= -- to obtain the desired control flow and degree of synchronization.
A basic example of non-blocking exchange of data between two processes is shown in Listing~\ref{lst:exchange}.

Note that this section introduces the API of \verb=numba-mpi= as in the most recent v1.0 release at the time of writing this paper.
For a more detailed and continuously updated description, please refer to the auto-generated docstring-based documentation hosted at \url{https://numba-mpi.github.io/numba-mpi}. 

\subsection{numba-mpi package availability}

Thanks to relying on Python built-in \verb=ctypes= foreign function interface, the package is implemented entirely in Python and thus \verb=numba-mpi= ``wheels'' (package files) are cross-platform.
The run-time dependencies of \verb=numba-mpi= are \verb=numba=, \verb=numpy=, and \verb=mpi4py= (plus \verb=pytest= if intending to run the tests).

Packages are available on PyPI (\url{https://pypi.org/p/numba-mpi}),\\ Conda Forge (\url{https://anaconda.org/conda-forge/numba-mpi}) and\\ AUR (\url{https://aur.archlinux.org/packages/python-numba-mpi}). 

\subsection{Test suite and workflows}

A suite of unit tests constitutes roughly half of the \verb=numba-mpi= codebase.
Test execution is automated using the \verb=pytest= framework.
The test suite, which covers the exposed functionality entirely, serves several purposes:
\begin{itemize}
 \item to check that data is transferred correctly between the wrappers and the underlying API;
 \item to ensure that the \texttt{numba-mpi} wrappers correctly map to underlying MPI functions;
 \item to verify that the wrappers function work correctly within Numba JIT-compiled functions as well as when executed from plain Python;
 \item to check compatibility with different MPI implementations under different operating systems.
\end{itemize}

\begin{Listing}[th!]
\begin{lstlisting}
import numpy as np
import pytest
import numba_mpi as mpi
from tests.common import data_types
from tests.utils import get_random_array

@pytest.mark.parametrize("snd, rcv", (
    (mpi.send, mpi.recv),
    (mpi.send.py_func, mpi.recv.py_func))
)
@pytest.mark.parametrize("fortran_order", [True, False])
@pytest.mark.parametrize("data_type", data_types)
def test_send_recv(snd, rcv, fortran_order, data_type):
    src = get_random_array((3, 3), data_type)
    if fortran_order:
        src = np.asfortranarray(src, dtype=data_type)
    else:
        src = src.astype(dtype=data_type)
    dst_exp = np.empty_like(src)
    dst_tst = np.empty_like(src)

    if mpi.rank() == 0:
        status = snd(src, dest=1, tag=11)
        assert status == mpi.SUCCESS
        COMM_WORLD.Send(src, dest=1, tag=22)
    elif mpi.rank() == 1:
        status = rcv(dst_tst, source=0, tag=11)
        assert status == mpi.SUCCESS
        COMM_WORLD.Recv(dst_exp, source=0, tag=22)
        np.testing.assert_equal(dst_tst, src)
        np.testing.assert_equal(dst_exp, src)
\end{lstlisting}
  \caption{\label{lst:test}An example test case that focuses on validating \texttt{numba-mpi} wrappers for MPI \texttt{Send} and \texttt{Recv} routines.}
\end{Listing}

Listing~\ref{lst:test} depicts one of unit tests for send/recv functionality that verifies handling of row-major and column-major dimension order.

The testing pipeline is executed via GitHub Actions: (i) whenever a pull request is created or updated; (ii) whenever any new commit arrives in the \verb=main= branch; and (iii) in weekly intervals to detect any incompatibilities with external packages.
Upon tagging a new release, the testing pipeline is extended to include upload of packaged "wheels" to the pypi.org package repository (while each \verb=main= branch commit triggers an upload to \verb=test.pypi.org=).

As of time of writing, testing is conducted on three different GitHub Actions runner images using: macOS 12.7.5, Ubuntu 22.04 and Windows Server 2022; all using x86\_64 hardware.
Tests are run without relying on network communication within MPI -- only inter-process communication is employed.
Table~\ref{tab:mpi_impls_and_OSes} lists different MPI implementations tested. That process is streamlined thanks to the \verb=setup-mpi= GitHub Action maintained by the \verb=mpi4py= project team.
Each configuration undergoes testing across five Python versions: 3.8, 3.9, 3.10, 3.11, and 3.12.

\begin{table}[tb]
\caption{\label{tab:mpi_impls_and_OSes}
    Operating systems and MPI implementations supported during the testing workflow as of time of writing. 
    Note: For configuration marked with "*", tests for \texttt{MPICH} on Linux image and Python $<3.10$ are disabled due to an external incompatibility.
}
\centering
\begin{tabular}{c|c|c|c|c}
\multicolumn{1}{c|}{\raisebox{-1.5ex}[0pt][0pt]{OS}} & \multicolumn{4}{c}{MPI implementation} \\ \cline{2-5}
\multicolumn{1}{c|}{} & \texttt{MPICH} & \texttt{OpenMPI} & \texttt{MS MPI} & \texttt{IntelMPI} \\ \hline
\texttt{Linux} & \hphantom{*}+* & + & - & + \\
\texttt{macOS} & + & + & - & - \\
\texttt{Windows} & - & - & + & - 
\end{tabular}
\end{table}

Additionally, \texttt{numba-mpi} incorporates automated workflows checking adherence to best practices in terms of code readability and maintainability. 
These tools include \verb=pylint=, \verb=black= and \verb=isort=.
A separate workflow ensures that the example code snippets provided in the \verb=README= file remain functional (including execution of the performance analysis presented in Fig.~\ref{fig:perf}).

\section{Illustrative examples of use in external software}\label{sec:examples}

\subsection{Usage of \texttt{numba-mpi} in \texttt{py-pde}}

The Python package \texttt{py-pde}~\cite{Zwicker_2020} allows numerical solutions of PDEs, which often appear when dynamical systems are analyzed.
The package focuses on PDEs defined on simple geometries, like rectangular, cylindrical, or spherical regions. 
Fields defined on those regions are represented at uniformly placed grid points, and the associated differential operators are expressed using finite differences.
In all cases supported by \texttt{py-pde}, this discretization allows to determine the values of differential operators based on the values at the support point and the neighboring points on the grid.
To provide a suitable definition for boundary points, \texttt{py-pde} extends the grid by virtual points, whose value is determined based on boundary conditions.
Since space is fully discretized, the PDEs are converted to a set of coupled ordinary differential equations, which can be solved using standard methods.
This approach provides a general method for solving  time-dependent PDEs coupling multiple fields of various ranks.

The parallelization of the approach taken by \texttt{py-pde} is fairly straight-forward since the grids can be decomposed along the orthogonal coordinate lines.
The main idea is that each MPI node is responsible for a subpart of the grid, and evolves the full equation analogously to a serial program.
The communication between nodes takes place via special boundary conditions, where the values of virtual points on one subgrid are dictated by the actual values of the respective adjacent grids.
Consequently, all sub-grids first exchange the values of their boundary points using \texttt{mpi\_send} and \texttt{mpi\_recv} and then advance their state according to a common time step.
The root MPI node is additionally responsible for data input and output, as well as adapting the time step if adaptive stepping is used.
This whole machinery is implemented within \texttt{py-pde}, so that the user does not need to set any of the details explicitly.

\begin{Listing}[t]
\begin{lstlisting}
import pde

grid = pde.UnitGrid([512, 512])
state = pde.ScalarField.random_uniform(grid, 0.49, 0.51)
eq = pde.PDE({"c": "laplace(c**3-c-laplace(c))-0.01*(c-0.5)"})

final_state = eq.solve(
    state,
    t_range=1e4,
    adaptive=True,
    solver="explicit_mpi",
    decomposition=[2, -1],
)
\end{lstlisting}
  \caption{\label{lst:py-pde}Example for using \texttt{py-pde} to solve equation~(\ref{eqn:ch}) for $k=10^{-2}$ and $c_0=\frac12$  with MPI support realized using \texttt{numba-mpi}.}
\end{Listing}

\begin{figure*}
  \centering
  \includegraphics[width=0.65\textwidth]{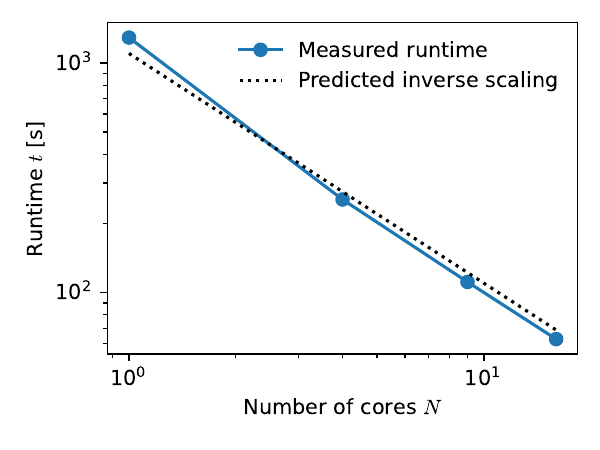}
  \caption{ 
	Runtime $t$ of  core calculation in Listing~\ref{lst:py-pde} as a function of the number $N$ of MPI cores.
	The predicted scaling $t \propto N^{-1}$ is indicated by the dotted line.
	Standard deviations determined from three repeated runs of the runtimes are smaller than the symbol size.
  }
  \label{fig:py-pde} 
\end{figure*}

As an example, Listing~\ref{lst:py-pde} shows a simple implementation of a simulation of the Cahn-Hilliard equation extended by chemical reactions, given by
\begin{equation}
	\label{eqn:ch}
	\partial_t c = \nabla^2\bigl(
		c^3 - c - \nabla^2 c
	\bigr) - k(c - c_0)
	\;.
\end{equation}
This equation describes the quick formation of droplets, which then slowly coarsen until they reach a maximal size controlled by the chemical reactions~\cite{Zwicker2015,Zwicker2022a}.
Analyzing these systems often requires large system sizes and long simulation times to gather adequate statistics.
Figure~\ref{fig:py-pde} shows that the runtime $t$ decreases dramatically with a larger number~$N$ of cores. 
The fact that $t$ decreases even faster than the predicted scaling $t\propto N^{-1}$ suggests that an increased computing efficiency, e.g., because the data now fits into the CPU cache, outperforms the overhead of the MPI communication.
With the help of \texttt{numba-mpi}, the \texttt{py-pde} package can provide efficiently parallelized numerical implementations of quite general non-linear partial differential equations.

\subsection{Usage of \texttt{numba-mpi} in \texttt{PyMPDATA-MPI}}

\begin{figure}[th!]
  \begin{subfigure}{0.495\textwidth}
    \centering
    \begin{adjustbox}{clip=true,trim=2cm .5cm 1.5cm 1cm}
        \includegraphics[width=1.35\linewidth]{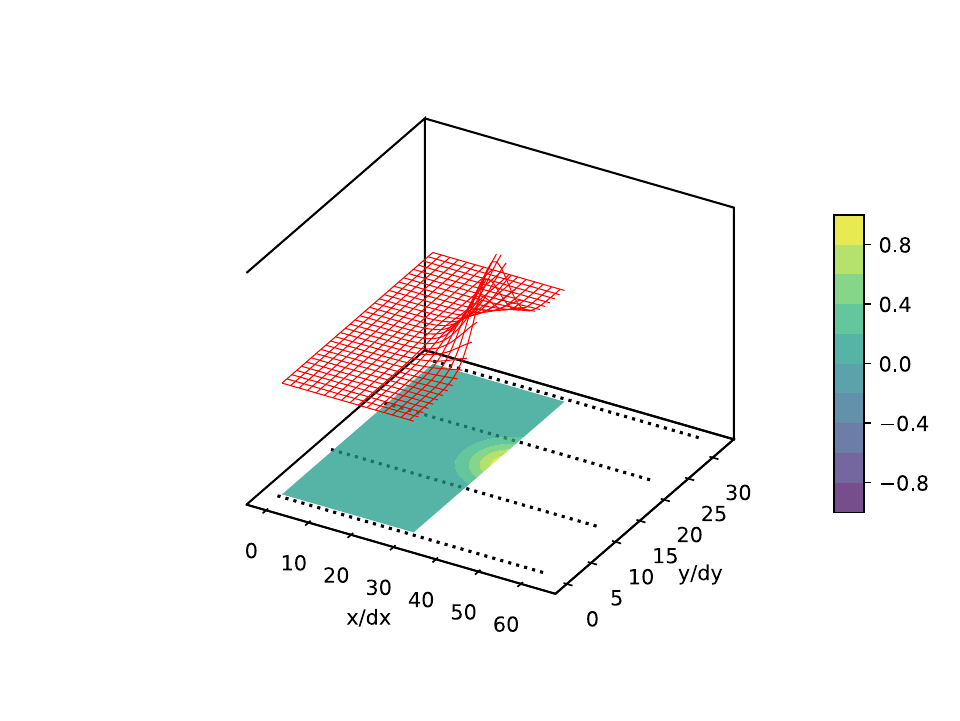} 
    \end{adjustbox}
    \caption{X: MPI (rank=0), Y: multi-threading}
  \end{subfigure}
  \begin{subfigure}{0.495\textwidth}
    \centering
    \begin{adjustbox}{clip=true,trim=2.2cm .5cm 0cm 1cm}
        \includegraphics[width=1.35\linewidth]{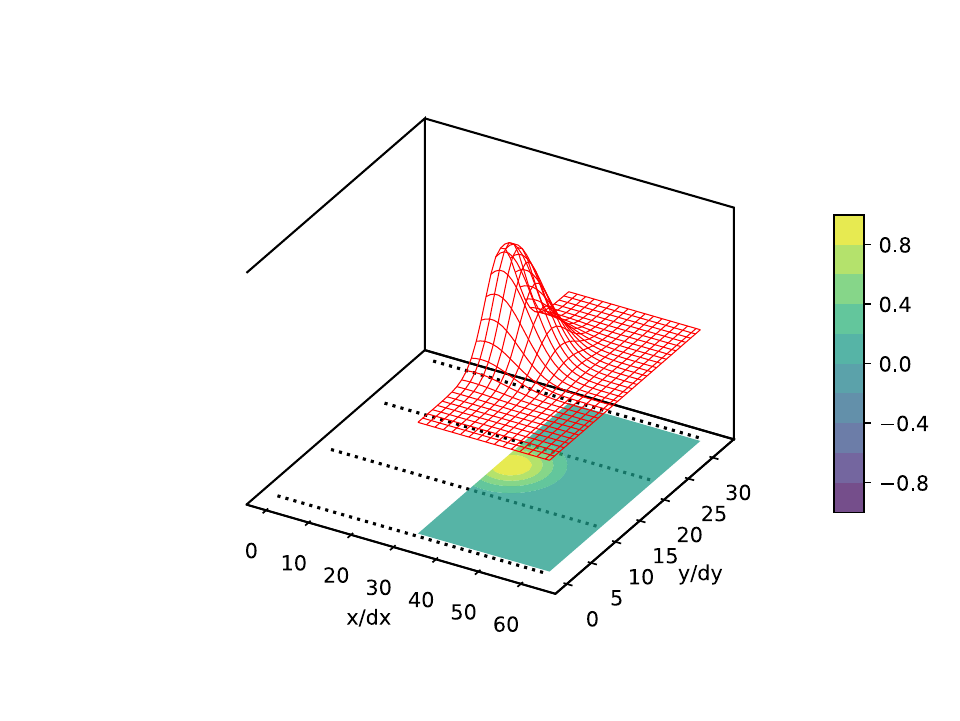}
    \end{adjustbox}
    \caption{X: MPI (rank=1), Y: multi-threading}
  \end{subfigure}
  \begin{subfigure}{0.495\textwidth}
    \centering
    \begin{adjustbox}{clip=true,trim=2cm .5cm 1.5cm .5cm}
        \includegraphics[width=1.35\linewidth]{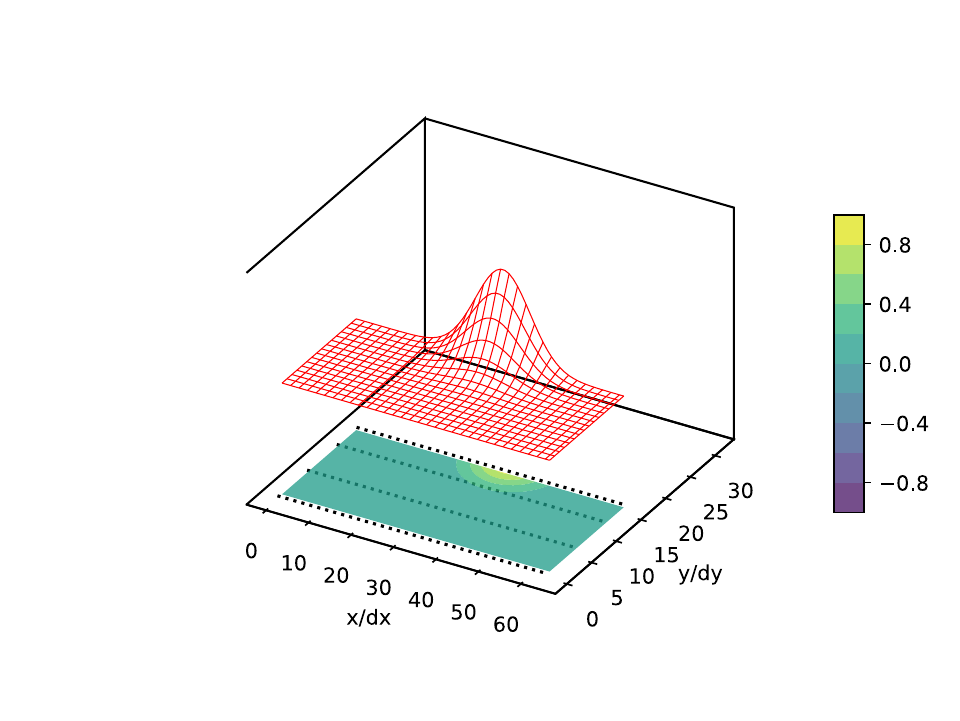} 
    \end{adjustbox}
    \caption{X: multi-threading \& MPI in X (rank=0)}
  \end{subfigure}
  \begin{subfigure}{0.495\textwidth}
    \centering
    \begin{adjustbox}{clip=true,trim=2.2cm .5cm 0cm .5cm}
        \includegraphics[width=1.35\linewidth]{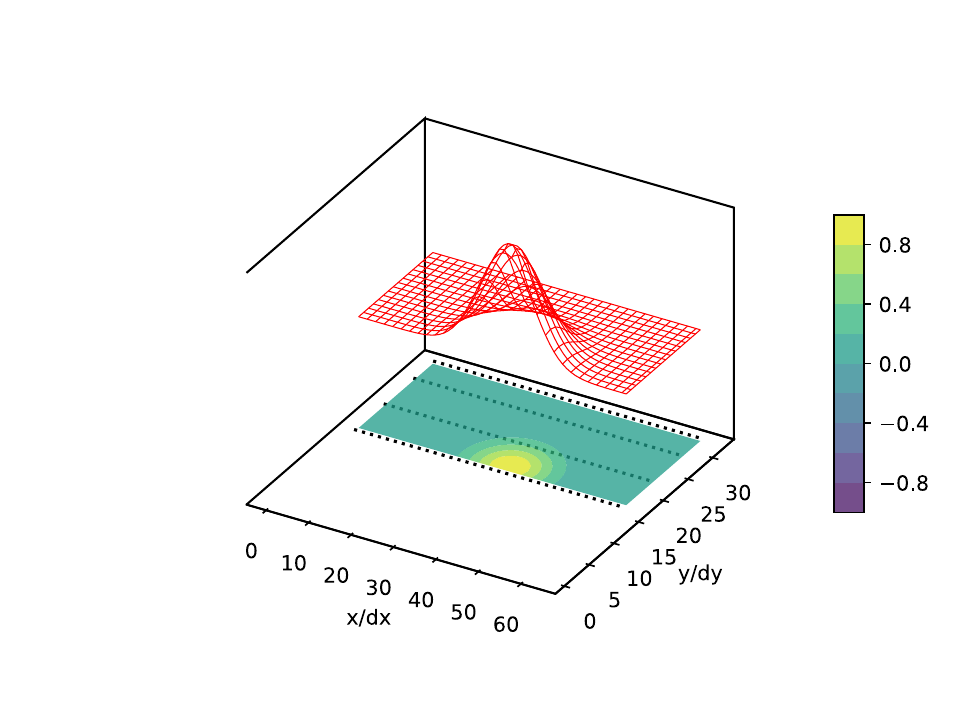}
    \end{adjustbox}
    \caption{X: multi-threading \& MPI in X (rank=1)}
  \end{subfigure}

  \caption{Different domain decomposition layouts tested in \texttt{PyMPDATA-MPI} with multi-threading (3 threads in all cases, dotted lines) and multi-processing (2 processes) carried out either along the same or distinct dimensions. The simulation setup involves a "hello-world" homogeneous advection problem with periodic boundary conditions.}
  \label{fig:pympdata}
\end{figure}

Handling MPI communication for domain decomposition is also done using \verb=numba-mpi= in the \verb=PyMPDATA-MPI=\footnote{\texttt{PyMPDATA-MPI} Python package: \url{https://pypi.org/p/PyMPDATA-MPI}} Python package, which provides a distributed memory extension to the \verb=PyMPDATA= \citep{Bartman_et_al_2022} suite of solvers for advection-diffusion PDEs given by 
\begin{equation}
    \partial_t (G \psi) + \nabla \cdot (G\vec{u} \psi) + \mu \nabla^2 (G \psi) = 0
    \;,
\end{equation}
where $\psi$ is a scalar field (advectee), $\vec{u}$ is a vector field (advector), $\mu$ is a diffusion coefficient, and $G$ corresponds to an optional coordinate transformation.
The solvers implement the Multidimensional Positive Definite Advection Transport Algorithm (MPDATA, see e.g. \cite{Waruszewski_et_al_2018} for an overview and a recently developed third-order accurate variant) -- an iterative finite-difference numerical scheme originating from the geophysical fluid dynamics community.
\verb=PyMPDATA= is engineered essentially entirely in Numba with both the algorithm iterations as well as the time-stepping loop embedded in JIT-compiled code blocks, what precludes usage of \verb=mpi4py= for domain decomposition.

There are three notable features of the Python-Numba-MPI solution stack, benefiting from \verb=numba-mpi=, exemplified in \verb=PyMPDATA-MPI=, namely:
\begin{itemize}
  \item possibility to implement hybrid multi-threading (shared-memory) and distributed parallelism in pure Python code thanks to OpenMP-based parallelization offered in Numba (multiple threads within JIT-compiled blocks of code are performing MPI communication simultaneously, using MPI tags to route the data to appropriate threads); 
  \item interoperability with MPI-IO used for handling simultaneous output from multiple processes to a single HDF5 file using the \verb=h5py= Python package \cite{Colette_et_al_2013} -- due to object-oriented interface of \verb=h5py=, the I/O logic needs to be outside of JIT-compiled blocks;
  \item flexibility in terms of parallelization strategy enabling choice of the dimension over which multi-threading and multi-processing are employed in domain decomposition to be done from user scope (see Fig~\ref{fig:pympdata}).
\end{itemize}

Altogether, this example highlights the strength of Python in terms of interoperability and its role as a "glue" language.
At the same time, as in the case of \verb=py-pde=, it highlights the strength of Numba as one of solution to the two-language problem \cite{Perkel_2019,Yang_et_al_2024} (i.e., the distinct technologies used for prototyping and high-performance applications).
It is \verb=numba-mpi= that enables the solution to also apply to distributed supercomputing applications. 
Notably, \verb=numba-mpi= not only serves to improve performance, but also 
  opens up the possibility to implement hybrid shared- and distributed-memory parallelism using threads and MPI.  

\section{Summary and outlook}\label{sec:concl}

Interoperability and a rich package ecosystem are among the key virtues of Python.
Numba is one of the leading technologies enabling high-performance computing with Python.
The MPI communication model has been available for use in Python code with the \verb=mpi4py= package, which however is not compatible with Numba.
The herein introduced \verb=numba-mpi= package, on the one hand, follows suite of developments such as \verb=numba-scipy=\footnote{\texttt{numba-scipy} Python package: \url{https://pypi.org/p/numba-scipy}}, \verb=numba-stats=\footnote{\texttt{numba-stats} Python package: \url{https://pypi.org/p/numba-stats}} and \verb=numba-kdtree=\footnote{\texttt{numba-kdtree} Python package: \url{https://pypi.org/p/numba-kdtree}} in offering Numba-compatible interfaces to packages relevant for scientific computing. 
On the other hand, \verb=numba-mpi= complements the wide range of packages offering access to the MPI API from languages other than C and Fortran (see \cite{MPI_hold} for discussion), including, e.g., the recently developed \verb=MPI.jl=\footnote{\texttt{MPI.jl} Julia package: \url{https://juliaparallel.org/MPI.jl}} Julia bindings \cite{Byrne_et_al_2021}, \verb=rsmpi= Rust bindings\footnote{\texttt{rsmpi} Rust package: \url{https://crates.io/crates/mpi}} \cite{Tronge_et_al_2023} and \verb=mpi4jax=\footnote{\texttt{mpi4jax} Python package: \url{https://pypi.org/p/mpi4jax}} \cite{Haefner_and_Vicentini_2021} which enables usage of MPI from JAX -- another JIT-compiler for Python.

In this paper, the rationale behind developing \verb=numba-mpi= was described by exemplifying (i) a two-order-of-magnitude execution-time speedup achievable with Numba JIT compilation; (ii) a multi-fold speedup achievable with embedding the entirety of MPI communications within JIT-compiled code blocks, which is not feasible with \verb=mpi4py=.
While the actual speedup figures will depend on the algorithms and data in question, the paramount importance of performance for typical MPI applications renders the cost of the ``roundtrips'' between JIT-compiled and Python code blocks worth profiling and potentially eliminating using \verb=numba-mpi=.

The reported performance gains and the existence of external software packages depending on \verb=numba-mpi= for handling domain-decomposition-related communication confirm the feasibility of engineering hybrid threading+MPI high-performance distributed computing systems in the Python-Numba-MPI technological stack.
Embracement of Numba, as compared to other Python acceleration technologies such as Cython \cite{Behnel_et_al_2011}, has the advantage of retaining familiar pure-Python codebase and the ability to fully leverage the debugging, development and maintenance tools from the Python ecosystem (when disabling JIT, which hampers profiling capabilities).

The \verb=numba-mpi= development is community-driven. 
The breadth of MPI API makes users' feedback all the more important for 
planning and prioritizing feature implementation. 
Work is under way on features including non-default communicators and wider support for non-contiguous array slices.
The public open-source development model, concise wrapper-type codebase and high test coverage help in lowering the entry-threshold for new developers. 
We encourage reporting feedback and engaging in code contributions.

\section*{Code availability}

Project development is hosted at \url{https://github.com/numba-mpi};\\
\verb=numba-mpi= and all its dependencies are free/libre and open-source software.
Releases of the package are persistently archived at Zenodo: \url{https://zenodo.org/doi/10.5281/zenodo.7336089}.
All listings presented in the paper are part of \verb=numba-mpi= v1.0.0 test suite.

\section*{Author contributions}

SA started development of \verb=numba-mpi= back in 2020 with a proof-of-concept interface to \verb=rank=/\verb=size=/\verb=initialized= and later led the project maintenance, contributor onboarding, packaging and dissemination activities. MM contributed implementation of \verb=send=/\verb=recv=, datatype handling and non-blocking communication with \verb=isend=/\verb=irecv=, \verb=wait[all|any]= and \verb=test[all|any]=, as well as interface to \verb=wtime= and the project logo. DZ contributed support for non-contiguous arrays, complex type, Fortran-ordered arrays, and reductions. KD contributed wrappers for \verb=bcast=, \verb=barrier= and \verb=gather=/\verb=scatter= operations and carried out performance/scalability tests of the library on the Cyfronet Ares supercomputer. OB contributed compatibility improvements for different MPI implementations and helped with project publicity. All co-authors further contributed to the project test suite and project maintenance. KD leads the development of \verb=PyMPDATA-MPI=, the relevant subsection of the paper is partly based on KD's MSc thesis carried out under supervision of SA.
DZ leads the development of \verb=py-pde= and authored the relevant subsection in the paper.
The paper text and code listings were drafted by SA \& KD with further input and feedback from all co-authors.

\section*{Conflict of Interest}

We wish to confirm that there are no known conflicts of interest associated with this publication and there has been no significant financial support for this work that could have influenced its outcome.

\section*{Acknowledgements}

We would like to thank: Numba Developers for help in sorting out the void-pointer challenges in the wrapping code; Piotr Bartman for participation in brainstorming at the early stage of the development; as well as all users who contributed feedback to the project through GitHub issues.

We gratefully acknowledge funding from the Polish National Science Centre (grant no. 2020/39/D/ST10/01220), the Max Planck Society and the European Union (ERC, EmulSim, 101044662).
We further acknowledge Poland’s high-performance computing infrastructure PLGrid (HPC Centers: ACK Cyfronet AGH) for providing computer facilities and support within computational grant no. PLG/2023/016369. 

\bibliographystyle{elsarticle-num}
\bibliography{refs}

\end{document}